\documentclass[twocolumn,aps,showpacs,prl,amsmath,amssymb,floatfix,nofootinbib]{revtex4-1}

\usepackage{color}
\usepackage{multirow}
\usepackage{bm} 
\usepackage{psfrag}
\usepackage[latin1]{inputenc}
\usepackage[english]{babel}
\usepackage{amsfonts}
\usepackage{amsmath}
\usepackage{upgreek} 
\usepackage[dvipsnames]{xcolor}

\usepackage{pdfpages} % include pdfs
\usepackage{pgffor} % for loops

\makeatletter
\AtBeginDocument{\let\LS@rot\@undefined}
\makeatother

\def\supplementfilename{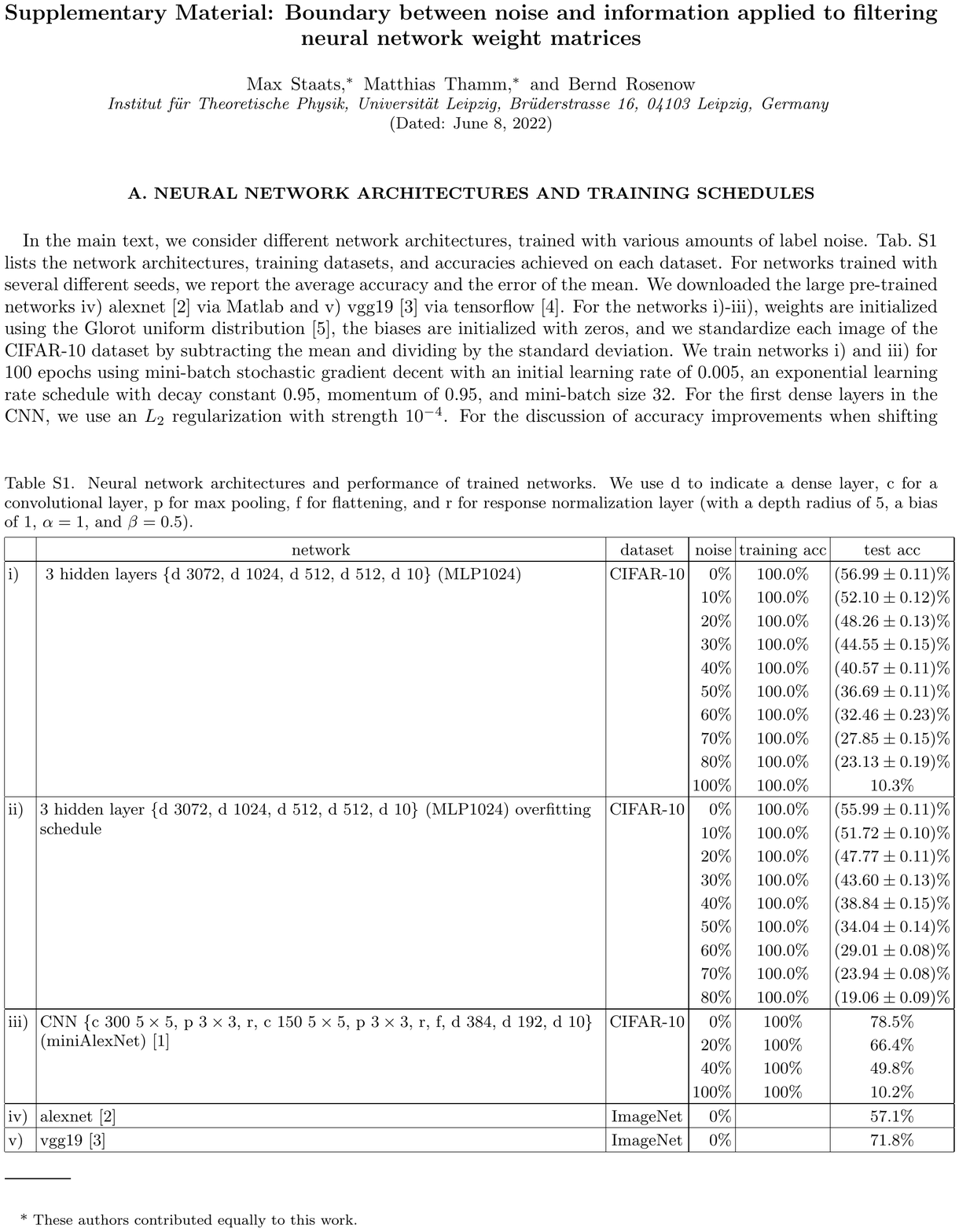}

\pdfximage{\supplementfilename}
\def\numbersupplementpages{\the\pdflastximagepages}

\newif\ifarXiv
\arXivtrue 
\DeclareSymbolFont{bbold}{U}{bbold}{m}{n}
\DeclareSymbolFontAlphabet{\mathbbold}{bbold}

\makeatletter
\def\blfootnote{\gdef\@thefnmark{}\@footnotetext}
\makeatother

\begin{document}

\title{
	Boundary between noise and information applied to filtering  neural network weight matrices
	}

\author{
	Max Staats
	}  
\thanks{These authors contributed equally to this work.}
\author{
  Matthias Thamm  
	} 
\thanks{These authors contributed equally to this work.}
\author{
	Bernd Rosenow
	}  
\affiliation{
	Institut f\"{u}r Theoretische Physik, Universit\"{a}t
  Leipzig,  Br\"{u}derstrasse 16, 04103 Leipzig, Germany
	} 
	
\date{\today}

\begin{abstract}
		Deep neural networks have been successfully applied to a broad range of problems where overparametrization yields weight matrices 
		which are partially random. A comparison of weight matrix singular vectors to the Porter-Thomas distribution suggests that there is a boundary between randomness and learned information in the singular value spectrum. Inspired by this finding, we introduce an algorithm for noise filtering, which both removes small singular values and reduces the magnitude of large singular values to counteract the effect of level repulsion between the noise and the information part of the spectrum. For networks trained in the presence of label noise, we indeed find that the generalization performance improves significantly due to noise filtering.  

\end{abstract}
 
\maketitle 
        
        \blfootnote{* These authors contributed equally to this work.} 
         \textit{Introduction: }
        In recent years, deep neural networks (DNNs) have proven to be powerful tools for solving a wide range of problems \cite{lecun2015,Goodfellow.2016,Carleo.2019,Silver.2017,bahri2020statistical}, including a large number of  applications in physics \cite{Carrasquilla.2017,vanNieuwenburg.2017,Ch.2017,Broecker.2017,Huembeli.2018,KoRi18,lee1997application,jin2021nsfnets,chen2021physics,duarte2018fast,guest2018deep}.
        Many of these networks are highly over-parametrized \cite{Goodfellow.2015,Soudry.2017,SiyuanMa.2018,Kawaguchi.2020,Simonyan.2014,Zhai.2021,cohen2021learning} and not only generalize well beyond the training data set, but also have the capacity to memorize large amounts of completely random  data \cite{lever2016points,Zhang.2021}.    However, if the training data set contains label noise, i.e.~mislabeled examples  (which is almost inevitable for realistic, large datasets \cite{frenay2013}),  overfitting the training data can significantly decrease the generalization performance \cite{Liu.2022}.

 In several studies, random matrix theory has been successfully applied to the analysis of neural networks \cite{Louart2018,Pennington2019,lampinen2019analytic,Pennington2018a,Baskerville.2021,granziol2020beyond,Martin.2021,Martin2021b,Thamm.2022}.
 Since weights of neural network are usually initialized randomly, learned information after training can be interpreted as deviations from random matrix theory (RMT) predictions, and it has been shown that even state of the art DNNs have weights that follow  RMT predictions   \cite{Martin.2021,Thamm.2022}. 
 The singular value distribution can be decomposed into a bulk and a tail region, where the random bulk can be fitted with a Marchenko-Pastur distribution
 appropriate for random matrices.  The theory of the Gaussian orthogonal ensemble accurately describes the level spacing and the level number variance of the bulk of the singular values in DNN weight matrices,   
 and most of the singular vectors follow the Porter-Thomas distribution. Deviations from RMT only occur  for  some large singular values and corresponding vectors \cite{Martin.2021,Thamm.2022}, explaining the success of low-rank approximations \cite{Xue.2013,Li.2018,Kim.2019,Liebenwein.2021,Alvarez.2017,Xu.2019, Idelbayev.2020_001}.

In this letter we study weight matrices for a variety of DNN models and architectures trained with and without label noise. 
We quantify how well  singular vector components follow the  Porter-Thomas distribution by 
applying a Kolmogorov-Smirnov test which takes into account the normalization of vectors, and find that there is a boundary between 
noise and information: 
singular vectors with small singular values  agree with the RMT prediction, while   vectors which large singular values significantly deviate. 
We confirm this finding  
by  setting singular values to zero, starting from the smallest one, and evaluate  the dependence of  the training and test accuracy on the percentage of removed singular values. We find that small singular values and the associated vectors indeed do not encode information. When training networks in the presence of label noise, more singular values are needed for a good training performance as compared to the pristine data set, indicating that label noise is  encoded predominately in intermediate singular values. Motivated by these results we suggest an algorithm for noise filtering of neural network weight matrices:  we filter the spectra from  noise by (i) removing small and intermediate  singular values and (ii) by reverting the shift of large singular values due to level repulsion with the noisy bulk. We find that this algorithm improves the test accuracy of DNNs trained with label noise by up to 6\;\%.

  {\em Setup:} We train several DNNs on the CIFAR-10 dataset \cite{Krizhevsky.2009} containing $N=50000$ training images $\bm{x}^{(k)}$ sorted into ten different classes, where the corresponding labels $\bm{y}^{(k)}$ are vectors with an entry of one at the position of the class and zeros otherwise. For training with label noise, we randomly shuffle a certain percentage of the labels  to force the networks to learn this noise in addition to the underlying rule. 
In a feed-forward DNN, the $l$-th layer  with $n_l$ neurons is described by an $n_{l-1}\times n_l$ weight  matrix ${\sf W}_l$, a bias vector $\bm{b}_l$ of length $n_l$, and an activation function $f_l$ such that an input image $\bm{x}$ activates the neurons in the input layer according to $\bm{a}_{0}  = \bm{x} $ and the activations are then passed through the network via
        \begin{align}
            \bm{a}_l = f_l({ \sf  W}_l \bm{a}_{l-1} + b_l)\ .
        \end{align}  
        The largest entry in the activation of the output layer $\bm{a}_{\rm out}$ then determines the network's prediction for the class of the input $\bm{x}$.
	\begin{figure}[tp]
		\centering
		\includegraphics[width=8.6cm]{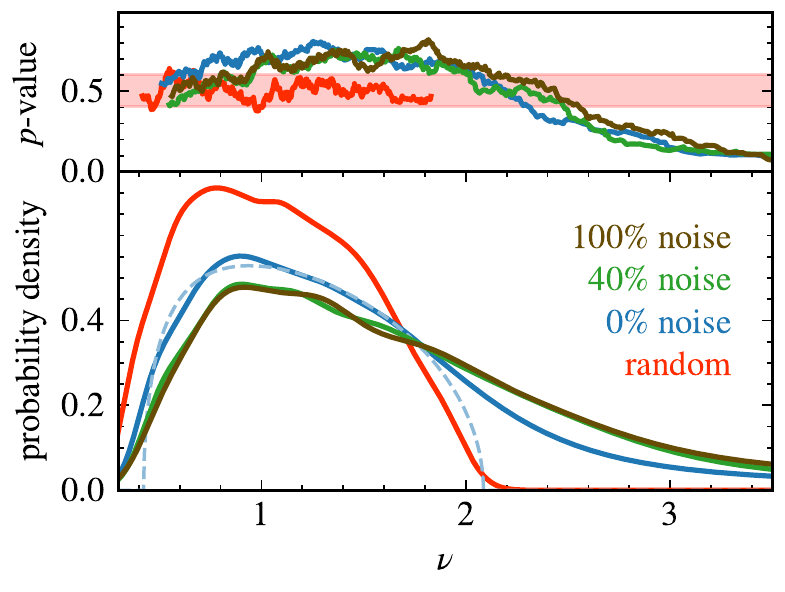}
		\caption{\label{Fig:mpspectraAndPvals} Analysis of singular values $\nu$ and vectors $V$ of the first hidden layer weight matrix for 
		the MLP1024 network trained  with various amounts of label noise: 0\% (blue), 40\% (green), and 100\% (brown). For reference, we show results for randomly initialized weights in red. The upper panel shows the randomness of singular vectors via the p-value of Kolmogorov-Smirnov tests against a Thomas-Porter distribution, averaged over neighboring singular values with a window size of 15; the light red stripe describes the $2\sigma$ region around the mean  for random vectors . The lower panel depicts the corresponding singular value spectra  obtained via Gaussian broadening with a window size of 15 (solid lines). The dashed line shows  the fit of a Marchenko-Pastur distribution to the spectrum for 0\% label noise. 
		}
	\end{figure} 
        For initialization of the networks, we choose zeros for the biases, draw random weights from a Glorot uniform distribution \cite{Glorot.2010}, and choose ReLU activations $f_l(\xi_i) = \max(\xi_i,0)$ for hidden layers and softmax $f_{\rm out}(\xi_i) = \exp({\xi_i})/\sum_{j} \exp({\xi_j})$ for the output layer.
        During training, the network is presented the image-label pairs in mini-batches of size 32 while gradient decent is used to adjust weights and biases such that the loss function
        \begin{align}
            \ell({\sf W},\bm{b}) &= -\frac{1}{N} \sum_{k=1}^N \bm{a}_{\rm out}^{(k)}\cdot \ln \bm{y}^{(k)}
        \end{align}
        is minimized, i.e.~such that the network's predictions agree with the labels (details of training parameters in \cite{Supplement}). 
        We train two kinds of architectures: (i) fully connected networks with  three hidden layers, denoted as MLP1024, with layer sizes [in, 1024, 512, 512, out], and (ii) modern convolutional neural networks (CNNs),  called miniAlexNet \cite{Zhang.2021}, consisting of two convolutional layers followed by max-pooling, batch normalization, and fully connected layers with regularization.
        To show that our results are generic, we additionally consider
        the two large networks alexnet \cite{Krizhevsky.2017} and vgg19 \cite{Simonyan.2014} pretrained on the  imagenet dataset \cite{Krizhevsky.2017} with 1000 classes. We   compute the singular value decompositions ${\sf W}={\sf U}\,\mathrm{diag}(\nu)\,{\sf V}$ with orthogonal matrices ${\sf U}$, ${\sf V}$ containing the singular vectors, and non-negative singular values $\nu$ for the dense fully connected layers described above. For convolutional layers in CNNs, we  first need to reshape the convolutional layer weight tensors to a rectangular shape and then perform the singular value decomposition described above \cite{Supplement}.           
  
\textit{Boundary between noise and information: }  	  
    For a random weight matrix in the limit of large matrix dimension, the  components of an  $m$-dimensional  singular vectors  follow a Porter-Thomas distribution, i.e.~a normal distribution with zero mean and standard deviation $1/\sqrt{m}$. 
    For the finite dimensional weight matrices we study here, the fact that singular vectors are normalized introduces correlations between the singular vector entries. 
    To account for these correlations, we compute the Kolmogorov-Smirnov test statistic with Monte-Carlo methods \cite{Supplement} and 
    apply the test to  the  empirical distribution of the singular vectors $V$ of trained networks.
    The resulting $p$-values, $0\leq p < 1$, indicate the likelihood of the sample to be from the test distribution.
    In order to reduce the amount of fluctuations in the $p$-values, we average over a window of size $15$ centered around  a given singular vector. The averaged $p$-values of a random control fluctuate around $0.5$ with a standard deviation of $\sigma = 0.05$.  

For the MLP1024 architecture in the absence of label noise, we find  that the averaged  $p$-values drop below two standard deviations of the random control for singular values larger than $2.3$, corresponding to $14.3$ percent of deviating singular values (blue solid line in the upper panel of Fig.~\ref{Fig:mpspectraAndPvals}), indicating that information may be contained in these these singular vectors. 
   For vectors corresponding to small singular values, the $p$-values lie within or even above the $2\sigma$ region  (this latter deviation is due to the requirement of orthogonality with vectors belonging to  large singular values \cite{Supplement}).  
Interestingly, the   $p$-values in the regime of large singular values  barely increase  in the presence of  40\% label noise. Only for completely random labels (100\% label noise) the $p$-values are somewhat increased, which is plausible since  random labels  should result in more random singular vectors. 
	
As a second approach \cite{Martin.2021,Thamm.2022}, we directly compare the empirical singular values to a Marchenko-Pastur distribution \cite{Marcenko.1967} valid for random matrices.  The bulk of small singular values can be fitted with a   Marchenko-Pastur distribution (lower panel), which describes the spectrum of randomly initialized weight matrices (dashed line, details in \cite{Supplement}). The upper end of such a 
fit is for singular values $\lambda \approx 2$, in agreement with the value $2.3$ found above for which the $p$-value is outside the 
$2 \sigma$-interval of the random control.  

	\begin{figure}[tp]
		\centering
		\includegraphics[width=8.6cm]{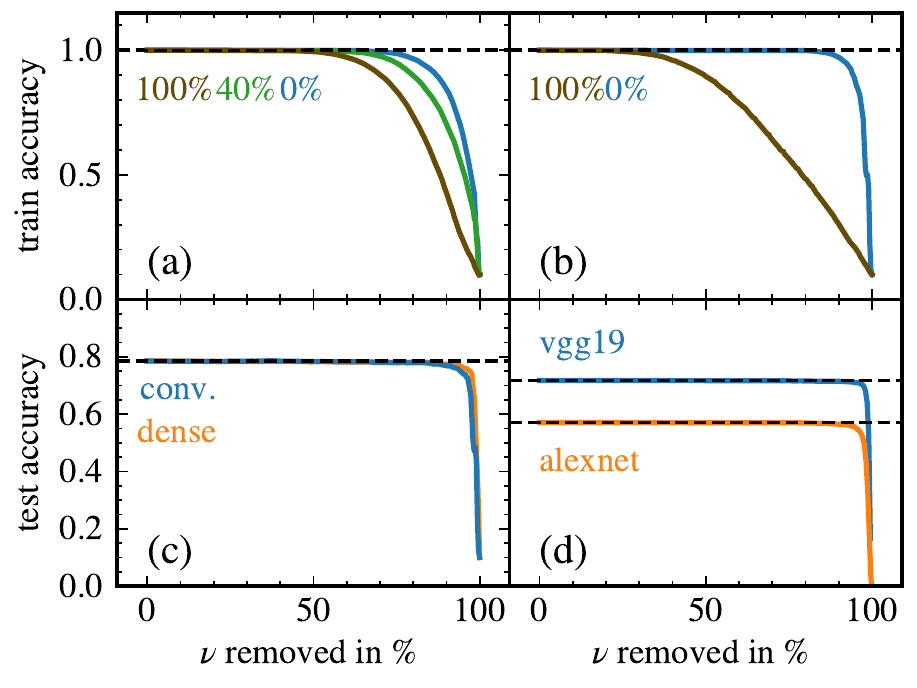}
		\caption{\label{Fig:removingSvals}  Information-noise
		boundary, demonstrated    by setting a given percentage of the singular values to zero.  (a) Training accuracy for the MLP1024 network  trained  with 
		various amounts of label noise (0\% blue, 40\% green, and 100\% brown).
		(b) Training accuracy for setting singular values from the second convolutional
		layer of miniAlexNet  trained  with 0\% label
		noise (blue) and 100\% label noise (brown) to zero.
		(c) Test accuracy for miniAlexNet trained without label noise for setting 
		singular values to zero in the first dense layer (orange) and the second 
		convolutional layer (blue).
		(d) Test accuracy for the pre-trained networks  vgg19 \cite{Simonyan.2014}
		(third dense layer, blue) and alexnet \cite{Krizhevsky.2017} (second dense layer, orange).
		In all cases, relevant information is stored in the largest singular values and 
		corresponding vectors only. In presence of label noise larger parts of 
		the spectrum are needed to store the noise.
		}
	\end{figure}
To verify the hypothesis  that system specific information is  stored in singular vectors corresponding to large singular values, we set singular values to zero, starting from the smallest one, 
and monitor the training and test accuracy of the networks (Fig.~\ref{Fig:removingSvals}). The training accuracy  (blue lines, MLP1024 in panel (a), miniAlexNet in panel (b)) indeed shows that learned information is stored only in the largest singular values and corresponding vectors.
	Label noise is learned differently from the rule and mainly stored in intermediate singular values such that the training accuracy drops significantly earlier in the case of label noise (40\% noise: green line, 100\% noise: brown lines; other amounts (not shown) interpolate). 
	This behavior is even more pronounced for a convolutional layer of miniAlexNet (Fig.~\ref{Fig:removingSvals}b), where the training accuracy drops sharply when removing more than 30\% of singular values, compared to a threshold of about 90\% for the pristine training data.

When considering the dependence of the test accuracy on the removal of singular values 
(Fig.~\ref{Fig:removingSvals}b and \ref{Fig:removingSvals}c)  it becomes apparent that generalization is solely due to the largest singular values and corresponding vectors.  
This is also valid for the large state of the art networks alexnet (orange) and vgg19 (blue) in Fig.~\ref{Fig:removingSvals}d. The observation that neural networks use only a small fraction of large singular values and vectors to learn the underlying rule sheds some light on the puzzle why neural networks generalize well even thought their capacity allows them to memorize random labels with ease, since larger singular values and vectors store the rule, and intermediate ones can be utilized for memorizing random labels. 
\begin{figure}[tp]
		\centering
		\includegraphics[width=8.6cm]{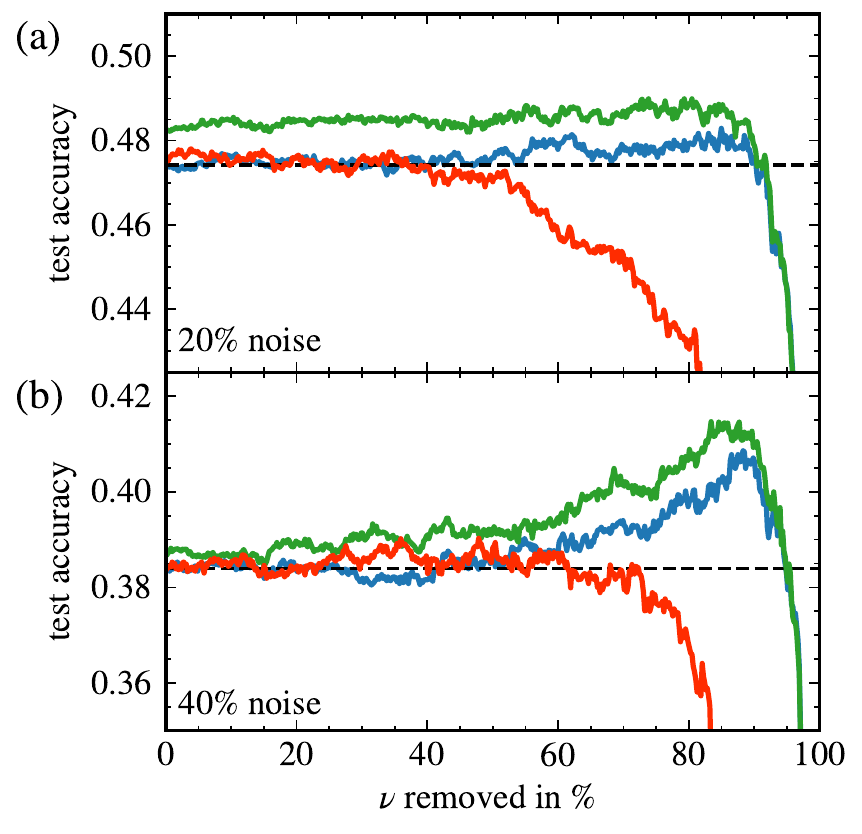}
		\caption{\label{Fig:improveEx}Dependence of the test accuracy on the removal and shifting of singular 
		values from the second hidden layer weights of MLP1024 networks trained in the presence of label noise: upon setting singular values to zero (blue) and when additionally shifting them according to Eq.~(3) (green) we observe a significant improvement in performance.  For training with overfitting (red) no improvement is observed, indicating that information and noise are mixed in the spectrum. }
	\end{figure} 

\textit{Noise filtering of weights: }    
   We next study how the generalization performance of DNNs trained with label noise depends 
    on the removal of singular values. In Fig.~\ref{Fig:improveEx} we show the impact on the test accuracy of setting  singular values to zero (blue lines)  in all hidden %in the first two hidden
    layers of MLP1024 networks trained with 20\% label noise (upper panel) and 40\% label noise (lower panel). In the latter case,  the generalization accuracy improves  by up to 2.5\% when removing about 90\% of singular values.	

	\begin{figure}[tp]
		\centering
		\includegraphics[width=8.6cm]{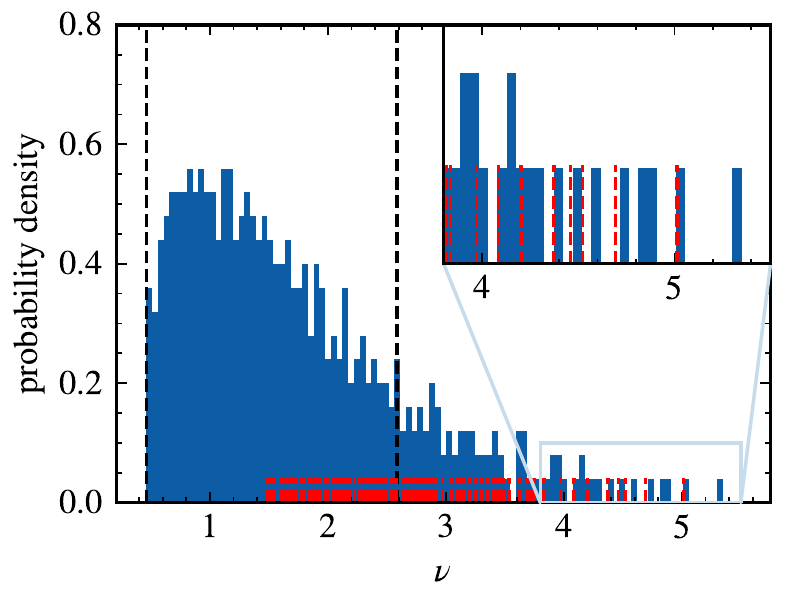}
		\caption{\label{Fig:shiftSvals}Shifting of singular values:  histogram of singular values   for the first hidden layer weight matrix of the
		MLP1024 network (blue) trained with 40\% label noise, with   boundaries of the 
		Marchenko-Pastur region  (dashed black lines).  
		The dashed red lines show the locations of shifted singular values  according to Eq.~\eqref{eq:rescale_sVal},   
		 and the inset zooms into the  tail region.
		}
	\end{figure} 

In addition,    we train another set of MLP1024 networks with severe overfitting, i.e.~we train for much longer than necessary to achieve 100\% training accuracy, with a slower learning rate schedule \cite{Supplement}. This causes an earlier drop of the test accuracy when removing singular values (red line, Fig.~\ref{Fig:improveEx}),  without any noticeable improvements before the drop. We interpret this behavior as a mixing between information and noise in the spectra such that no clear boundary between these regimes exists anymore. 

Level repulsion in random matrices  
can  lead to an upward shift of large singular values in the presence of a random bulk of smaller singular values. Such a shift of eigenvalues was discussed in \cite{Lampinen.2018} for the case of a linear network, and is also known to exist in empirical covariance matrices, e.g.~between returns of a stock portfolio \cite{Laloux.1999,Plerou.1999}.
For the latter case, estimators have  been developed which compensate for this singular value shift \cite{Bun2017,schaefer2010}.  Here,
we suggest to model  the contribution of the weights' noisy bulk ${\sf W}_{\rm noise}$ to the low rank  part ${\sf W}_0$ containing the information in an additive way, such that the weight matrix after training is given by ${\sf W}={\sf W}_0 + {\sf W}_{\rm noise}$. The upwards shift of large singular values due to level repulsion  can then be  explicitly computed \cite{benaych2012,Lampinen.2018} in the limit 
 where the dimensions of the  $n_{l-1}\times n_l$ weight matrix  tend to infinity with a fixed ratio $q=n_{l}/n_{l-1} \leq 1$. 
 Under the assumption of i.i.d~distributed elements of ${\sf W}_{\rm noise}$ with standard deviation $\sigma$, the singular values $\nu$ of ${\sf W}$ can be shifted back to recover the unperturbed singular values $\nu_0$ of ${\sf W}_0$ via   
	\begin{align}
	    \frac{\nu_0}{\sigma} = \sqrt{ \left(\frac{\nu}{\sigma}\right)^2 -q-1 +\sqrt{ \left( \left(\frac{\nu}{\sigma}\right)^2 -q-1\right)^2 -4q } } \ \ ,
	    \label{eq:rescale_sVal}
	\end{align}
where  the standard deviation of the noise term $\sigma$ is obtained from a Marchenko-Pastur fit to the spectrum (for details see \cite{Supplement}).
The effect of such a transformation is shown in Fig.~\ref{Fig:shiftSvals}: 
While large values are shifted by a relatively small amount as seen in the inset, there are several singular values that get pushed far into the MP region.  	
	\begin{figure}[tp] 
		\centering
		\includegraphics[width=8.6cm]{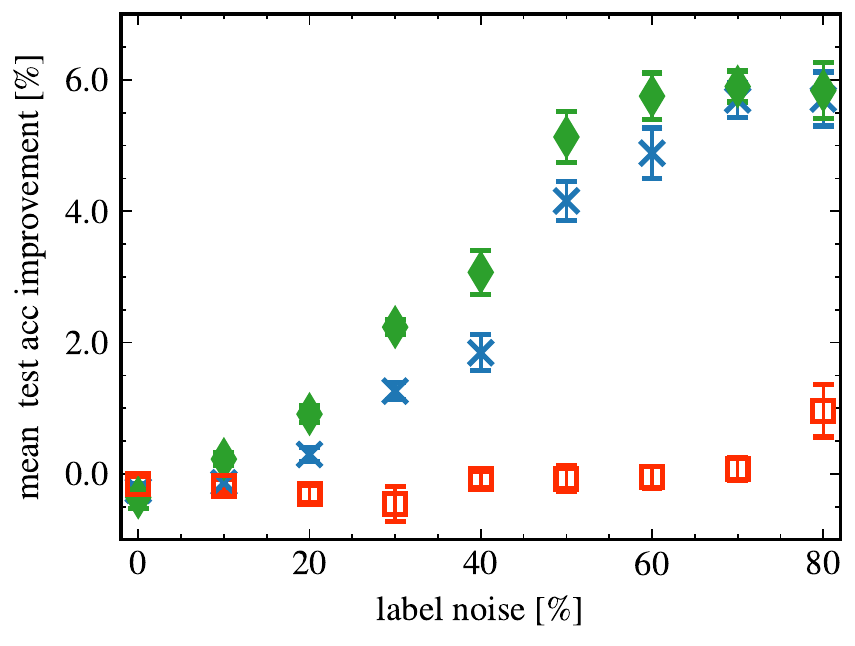}
		\caption{\label{Fig:meanImprovement} Average improvement of the test accuracy 
		when removing singular values (blue, red) from all layers and when additionally shifting singular 
		values (green) of the first two layers in MLP1024 networks, with results for both the  learning rate schedule 
		(blue crosses, green diamonds) and an overfitting schedule (red squares). 
		We observe that the average improvements increase with increasing amount
		of label noise, with an enhanced improvement for additionally shifting singular values. 
		There are no improvements for networks trained with overfitting.
				}
	\end{figure} 
When applying a shift of singular values originally outside the Marchenko-Pastur region together with the removal of singular values, we use the following algorithm: we i) rank order the singular values, then ii) shift the large singular values while keeping their rank within the spectrum unchanged, and then iii) remove singular values from small to large according to their rank. The green lines in 	Fig.~\ref{Fig:improveEx} describe the dependence of 
test accuracy on both shifting and removing singular values  in the first two hidden layers of MLP1024 networks trained with 20\% label noise (upper panel) and 40\% label noise (lower panel), respectively. We find significant improvements of the generalization accuracy using a combination of shifting and removing singular values (similar for the convolutional layer of miniAlexNet \cite{Supplement}) as compared to only 
removing them.
	
	To study the typical improvement in  test accuracy 
	when noise filtering DNN  weight matrices, we train 
	MLP1024 networks initialized using ten different seeds  on the 50000 CIFAR-10 training images for both training schedules at amounts of label noise varying between 0\% and 80\%. We split the remaining 10000 images from the dataset into a validation set of 2000 images and a test set of 8000 images. Using the validation data we determine the optimum amount of singular values to be removed for 
	all layers, except for the output ones with only ten singular values. We optimize by moving from the last layer to the first one, 
	while always fixing the previously filtered weights when optimizing the next layer. We determine the amount of singular values to be removed in each layer from the maximum of the validation accuracy, when consecutively setting singular values to zero, starting with the smallest one.
	In the cases where we additionally shift singular values, we compare the accuracy of both shifting and removing an optimal amount of singular values in a given layer layer 
	with the accuracy of only removing singular values, and then choose the algorithm with the better validation accuracy. In this process, 
	we do not optimize over the amount of singular values which are shifted, but instead determine the smallest singular value to be shifted by the upper boundary of a Marchenko-Pastur fit to the bulk part of the spectrum as described above.    
	
	In  Fig.~\ref{Fig:meanImprovement} one sees that 
	for  networks trained with the regular schedule   an average 
	increase of roughly 1\% already in the case of 20\%	label noise is achieved by a combination of shifting and removing (green symbols). Increasing the label noise further we find  significant 
	improvements of several percent. 
	However, for the overfitted 
	networks (red symbols)   
	no improvements of the test accuracy are obtained when filtering the 
	singular values. For the pretrained network vgg19 we find that shifting singular values of the last layers can  decreases the test accuracy,  while small improvements of $0.3\%$ percent  are possible on the validation data set when shifting and removing from the first two layers. However,  no improvements can be observed on the test set. Here, we randomly drew $2000$ images from the original imagenet test set as a validation set and another $8000$ as a our new test set such that the two sets are disjoint.

\textit{Conclusions.}
    By comparing singular vectors to the Porter-Thomas distribution and singular values to a Marchenko-Pastur law, we argue that weight matrices of DNNs exhibit a boundary between  noise  and information  in their spectra.
    We substantiate  this concept by  systematically setting singular values to zero while monitoring the impact on the training and test accuracy.	It turns out that (i) small singular values do not contribute to either training or test  performance, (ii) large singular values encode the underlying rule, and (iii) intermediate singular values are important for the training accuracy when learning images with label noise, even though 
   they correspond to random singular vectors. We suggest a filtering algorithm combining the removal of small singular values with the  downward shift of large ones and find that  it increases the generalization performance significantly in the presence of label noise.    
	As label noise can be inherent in datasets where label annotation is difficult, we believe that filtering of weight matrices could be useful for improving the performance of DNNs in such situations. 
	\newline 
	
\begin{acknowledgments}  
	 \textit{Acknowledgments.} This work has been funded by the Deutsche Forschungsgemeinschaft (DFG) 
	 under Grants No.~RO 2247/11-1 and No.~406116891 within the Research 
	Training Group RTG 2522/1.\\ 
\end{acknowledgments}

%\bibliography{bibliography} 
%apsrev4-2.bst 2019-01-14 (MD) hand-edited version of apsrev4-1.bst
%Control: key (0)
%Control: author (8) initials jnrlst
%Control: editor formatted (1) identically to author
%Control: production of article title (0) allowed
%Control: page (0) single
%Control: year (1) truncated
%Control: production of eprint (0) enabled
%

\ifarXiv
    

\section*{Supplementary material}

# Supplementary Material: Boundary between noise and information applied to filtering neural network weight matrices

Max Staats,* Matthias Thamm,* and Bernd Rosenow

*Institut für Theoretische Physik, Universität Leipzig, Brüderstrasse 16, 04103 Leipzig, Germany*

(Dated: June 8, 2022)

## A. NEURAL NETWORK ARCHITECTURES AND TRAINING SCHEDULES

In the main text, we consider different network architectures, trained with various amounts of label noise. Tab. S1 lists the network architectures, training datasets, and accuracies achieved on each dataset. For networks trained with several different seeds, we report the average accuracy and the error of the mean. We downloaded the large pre-trained networks iv) alexnet [2] via Matlab and v) vgg19 [3] via tensorflow [4]. For the networks i)-iii), weights are initialized using the Glorot uniform distribution [5], the biases are initialized with zeros, and we standardize each image of the CIFAR-10 dataset by subtracting the mean and dividing by the standard deviation. We train networks i) and iii) for 100 epochs using mini-batch stochastic gradient decent with an initial learning rate of 0.005, an exponential learning rate schedule with decay constant 0.95, momentum of 0.95, and mini-batch size 32. For the first dense layers in the CNN, we use an $L_2$ regularization with strength $10^{-4}$. For the discussion of accuracy improvements when shifting

Table S1. Neural network architectures and performance of trained networks. We use d to indicate a dense layer, c for a convolutional layer, p for max pooling, f for flattening, and r for response normalization layer (with a depth radius of 5, a bias of 1, $\alpha = 1$, and $\beta = 0.5$).

| | network | dataset | noise | training acc | test acc |
|---|---|---|---|---|---|
| i) | 3 hidden layers {d 3072, d 1024, d 512, d 512, d 10} (MLP1024) | CIFAR-10 | 0% | 100.0% | $(56.99 \pm 0.11)\%$ |
| | | | 10% | 100.0% | $(52.10 \pm 0.12)\%$ |
| | | | 20% | 100.0% | $(48.26 \pm 0.13)\%$ |
| | | | 30% | 100.0% | $(44.55 \pm 0.15)\%$ |
| | | | 40% | 100.0% | $(40.57 \pm 0.11)\%$ |
| | | | 50% | 100.0% | $(36.69 \pm 0.11)\%$ |
| | | | 60% | 100.0% | $(32.46 \pm 0.23)\%$ |
| | | | 70% | 100.0% | $(27.85 \pm 0.15)\%$ |
| | | | 80% | 100.0% | $(23.13 \pm 0.19)\%$ |
| | | | 100% | 100.0% | 10.3% |
| ii) | 3 hidden layer {d 3072, d 1024, d 512, d 512, d 10} (MLP1024) overfitting schedule | CIFAR-10 | 0% | 100.0% | $(55.99 \pm 0.11)\%$ |
| | | | 10% | 100.0% | $(51.72 \pm 0.10)\%$ |
| | | | 20% | 100.0% | $(47.77 \pm 0.11)\%$ |
| | | | 30% | 100.0% | $(43.60 \pm 0.13)\%$ |
| | | | 40% | 100.0% | $(38.84 \pm 0.15)\%$ |
| | | | 50% | 100.0% | $(34.04 \pm 0.14)\%$ |
| | | | 60% | 100.0% | $(29.01 \pm 0.08)\%$ |
| | | | 70% | 100.0% | $(23.94 \pm 0.08)\%$ |
| | | | 80% | 100.0% | $(19.06 \pm 0.09)\%$ |
| iii) | CNN {c 300 $5 \times 5$, p $3 \times 3$, r, c 150 $5 \times 5$, p $3 \times 3$, r, f, d 384, d 192, d 10} (miniAlexNet) [1] | CIFAR-10 | 0% | 100% | 78.5% |
| | | | 20% | 100% | 66.4% |
| | | | 40% | 100% | 49.8% |
| | | | 100% | 100% | 10.2% |
| iv) | alexnet [2] | ImageNet | 0% | | 57.1% |
| v) | vgg19 [3] | ImageNet | 0% | | 71.8% |

_____

* These authors contributed equally to this work.

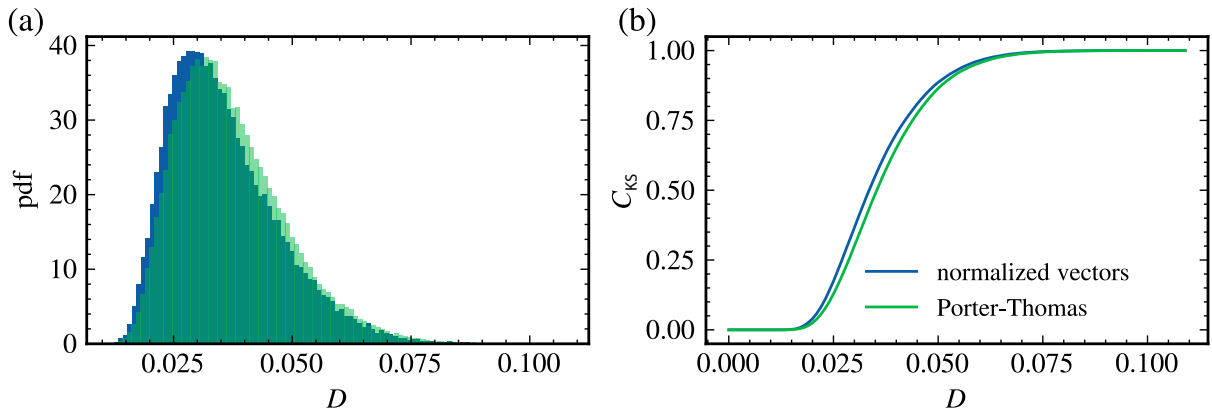

Figure S1. Comparison between Kolmogorov-Smirnov test statistics for random vectors of length 512 with i.i.d. entries (green) and for normalized vectors (blue). (a) Probability density function of the Kolmogorov-Smirnov distances Eq. (S4) obtained with Monte-Carlo sampling. (b) Cumulative distribution functions for the pdfs shown in (a). It becomes apparent that normalizing the vectors significantly changes the Kolmogorov-Smirnov test statistics even though it leaves the cumulative distribution of the vector entries unchanged.

and removing singular values we also consider an overfitting training schedule ii) with 500 epochs, with a stepwise schedule starting at a learning rate 0.001, which is then reduced it by a factor of 0.7 every 50 epochs. This ensures that we train for a large number of epochs after reaching 100% training accuracy.

## B. RESHAPING OF CONVOLUTIONAL LAYER WEIGHTS

For convolutional layers, weights are 4D tensors such that we need to reshape them first to a matrix shape before computing the singular value decomposition. A weight $W$ of shape $(K, L, M, N)$ with $K > L > M > N$ is reshaped to a matrix $\tilde{W}$ with sizes $(K, L \cdot M \cdot N)$ according to

$$\tilde{W}_{k,(l \cdot M \cdot N + m \cdot N + n)} = W_{k,l,m,n} \tag{S1}$$

with indices counted from zero. This way, elements that are close to each other in the tensor remain close in the matrix [6].

## C. FITTING MARCHENKO-PASTUR CURVES

In the main text, we fit the singular value density of DNN weight matrices to the Marchenko-Pastur (MP) distribution [7]

$$P(\nu) = \begin{cases} \frac{n/m}{\pi \sigma^2 \nu} \sqrt{(\nu_{\max}^2 - \nu^2)(\nu^2 - \nu_{\min}^2)} & \nu \in [\nu_{\min}, \nu_{\max}] \\ 0 & \text{else} \end{cases} \tag{S2}$$

with $\nu_{\max \atop \min} = \sigma(1 \pm \sqrt{m/n})$ to obtain the standard deviation of the noise term $\sigma$ used in the shifting formula Eq. (3). As the spectrum additionally has singular values in the tail, the MP part is not normalized and the end of the MP region is not known a priori. We therefore first broaden the DNN spectrum using Gaussian broadening [8]

$$P(\nu) \approx \frac{1}{m} \sum_{k=1}^{m} \frac{1}{\sqrt{2\pi\sigma_k^2}} \exp\left(-\frac{(\nu - \nu_k)^2}{2\sigma_k^2}\right) \ . \tag{S3}$$

with $\sigma_k = (\nu_{k+a} - \nu_{k-a})/2$ and windows size $a = 15$, and then fit an adjusted MP distribution, where we use $\nu_{\max}$ and the maximum height as independent fit parameters, and infer $\nu_{\min}$ from the smallest singular values. This yields an estimate for $\nu_{\min}$ and $\nu_{\max}$. We then fit the proper MP distribution Eq. (S2), only depending on $\sigma$, to a normalized histogram of the singular values between $\nu_{\min}$ and $\nu_{\max}$.

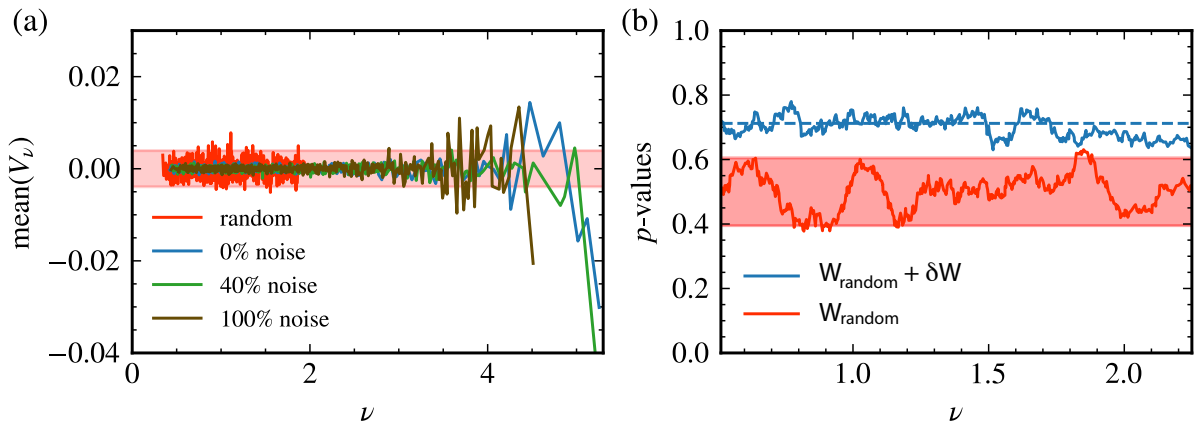

Figure S2. (a) Mean values of singular vector components for the first hidden layer of the trained MLP1024 DNN (same vectors as in Fig. 1 main text) as a function of the corresponding singular values $\nu$. The red line shows the distribution of means for singular vectors of a random random weight matrix with i.i.d. Gaussian entries, with the corresponding $2\sigma$ region shown as transparent red stripe. We observe means much closer to zero for singular vectors of the trained weight matrix in the case of small singular values, and significantly larger means for the vectors corresponding to large singular values. (b) Kolmogorov-Smirnov $p$-values (with test statistics from Sec. D) for singular vectors of a $1024 \times 512$ matrix $W_{\text{random}}$ (red) with i.i.d. Gaussian entries with zero mean and variance $1/512$; $2\sigma$ region for $p$-values shown in light red. When adding a matrix $\delta W$ of rank ten with entries from a Gaussian distribution with the same variance but with mean $-0.01$ (similar to mean values observed for empirical vectors corresponding to the largest singular values in (a)), the sum $W_{\text{random}} + \delta W$ (blue) has singular vectors with significantly increased $p$-values, due to the requirement of orthogonality between singular vectors with large and small singular values. The $p$-values are averaged over neighboring singular values with a window size of 15.

## D. KOLMOGOROV-SMIRNOV TEST STATISTIC FOR NORMALIZED PORTER-THOMAS VECTORS

Entries of $N$-dimensional singular vectors $\xi_i$ of random matrices from the Gaussian orthogonal ensemble follow the cumulative Porter-Thomas distribution function $C_{\text{PT}}(x) = 1/2 + \text{erf}\left(\sqrt{N/2}\,x\right)/2$. However, their entries are not uncorrelated due to the normalization condition $\sum_i \xi_i^2 = 1$. Hence, the statistic of the usual Kolmogorov-Smirnov test which determines the $p$-values for uncorrelated data cannot be applied here. We obtain the statistic for normalized vectors using Monte-Carlo sampling of 50000 normalized random vectors $\boldsymbol{\xi}^{(k)}$ by computing the empirical cdf for each vector $C_{\text{emp}}^{(k)}(\boldsymbol{\xi}^{(k)})$ to find the corresponding Kolmogorov-Smirnov distances

$$D^{(k)} = \sup_x |C_{\text{emp}}^{(k)}(\boldsymbol{\xi}^{(k)}) - C_{\text{PT}}(\boldsymbol{\xi}^{(k)})| \ . \tag{S4}$$

The cdf $C_{\text{KS}}(D)$ for the 50000 distances $\{D^{(k)}\}$ allows to determine the $p$-values for a given new vector $\boldsymbol{\xi}$ with deviation $D(\boldsymbol{\xi})$ as $1 - p = C_{\text{KS}}(D(\boldsymbol{\xi}))$. The deviations between the usual Kolmogorov-Smirnov statistic (green) and the sampled statistic for normalized vectors (blue) are shown in Fig. S1.

## E. INCREASED $p$-VALUES

In main text Fig. 1 we test the singular vector entries of trained weight matrices against the Porter-Thomas distribution and find that the $p$-values in the random part of the spectrum are significantly higher than statistically expected. We argue that this is due to the presence of a few non-random singular vectors that store the information. These vectors force the random singular vectors to have a narrower distribution around the most likely part of the Porter-Thomas distribution (normal distribution with zero mean) due to the constraint of orthogonality with the deviating singular vectors with large singular values.

For example, using the same test statistic as described in Sec. D such that random normalized vectors from the Porter-Thomas distribution have on average a $p$-value of 0.5, the subset of vectors with zero mean have an average $p$-value of 0.74. We show in Fig. S2(a) that the mean values of singular vector entries for small singular values of trained weight matrices (0% label noise blue, 40% green, 100% brown) are indeed smaller than the expected values

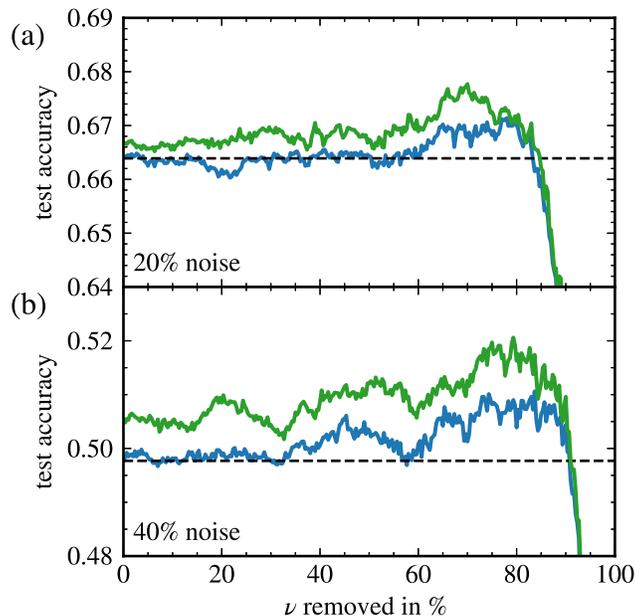

Figure S3. Dependence of the test accuracy on the removal and shifting of singular values from the weight matrix of the first convolutional layer in miniAlexNet after training with label noise. The dashed lines depict the test accuracy without removing singular values. We compare setting singular values to zero (blue) with additionally shifting the singular values (green) according to main text Eq. (3). (a) For training on CIFAR-10 with 20% label noise, and (b) for 40% label noise.

($2\sigma$ range in light red stripe) for fully random matrices (red) while the means are much larger for large singular values where the information is stored.

The increase of $p$-values can also be shown for a simple model, adding a low-rank matrix $\delta W$ to a fully random matrix $W_{random}$ that would have singular vectors with $p$-value of 0.5 on average. For this we draw a $1024 \times 512$ matrix $W_{random}$ with Gaussian distributed i.i.d. entries with mean zero and variance $1/512$, for which the $p$-values of singular vectors fluctuate around 0.5 (see red curve in Fig. S2(b), with most values within the $2\sigma$ region (light red stripe)). We then draw a second $1024 \times 512$ matrix with i.i.d. Gaussian distributed entries with mean $-0.01$ and variance $1/512$, compute the singular value decomposition, and reconstruct the matrix by only keeping the largest 10 singular values yielding a rank 10 matrix $\delta W$. We then analyze the $p$-values of the singular vectors of $W_{random} + \delta W$. We find that the $p$-values are increased (blue line in Fig. S2(b)), with mean 0.71, which shows that in the presence of a few singular vectors with a distribution different from the random bulk, we expect the $p$-values in the bulk to be increased due to the enforced orthogonality to the singular vectors with a finite mean.

## F. CLEANING OF A CONVOLUTIONAL LAYER

In Fig. S3 we show improvements of the test accuracy when filtering weights from convolutional layers of miniAlexNet networks (analogous to Fig. 4 in the main text), both with and without shifting of eigenvalues . The results are very similar to the behavior for dense layers described in the main text.

---


\fi

\end{document}